\documentclass{ws-ijmpa}
\usepackage[super,compress]{cite}
\usepackage{graphicx}
\usepackage{cancel}
\usepackage{cite}
\usepackage{slashed}
\usepackage{amsmath,bbm}
\usepackage{amssymb}
\usepackage{graphicx}
\usepackage[latin1]{inputenc}
\usepackage{mathrsfs}

\newcommand{\bea}{\begin{eqnarray}}
\newcommand{\eea}{\end{eqnarray}}
\newcommand{\be}{\begin{equation}}
\newcommand{\ee}{\end{equation}}
\newcommand{\ba}{\begin{array}}
\newcommand{\ea}{\end{array}}

\def\gsim{\mathrel{\rlap{\lower4pt\hbox{\hskip1pt$\sim$}}
    \raise1pt\hbox{$>$}}}

\begin{document}
\markboth{Stefan Antusch and Constantin Sluka}{Testable SUSY spectra from GUTs at a 100 TeV pp collider}

%%%%%%%%%%%%%%%%%%%%% Publisher's Area please ignore %%%%%%%%%%%%%%%
%
\catchline{}{}{}{}{}
%
%%%%%%%%%%%%%%%%%%%%%%%%%%%%%%%%%%%%%%%%%%%%%%%%%%%%%%%%%%%%%%%%%%%%

\title{Testable SUSY spectra from GUTs at a 100 TeV pp collider}

\author{Stefan Antusch\footnote{E-mail: stefan.antusch@unibas.ch} $^{1\,2}$ }

\author{Constantin Sluka\footnote{E-mail: constantin.sluka@unibas.ch} $^1$ }

\address{$^1$ Department of Physics, University of Basel,\\
Klingelbergstr. 82, CH-4056 Basel, Switzerland}

\address{$^2$ Max-Planck-Institut f\"ur Physik (Werner-Heisenberg-Institut),\\
F\"ohringer Ring 6, D-80805 M\"unchen, Germany}
\maketitle

%\begin{history}
%\received{Day Month Year}
%\revised{Day Month Year}
%\end{history}

\begin{abstract}
Grand Unified Theories (GUTs) are attractive candidates for more fundamental elementary particle theories. They can not only unify the Standard Model (SM) interactions but also different types of SM fermions, in particular quarks and leptons, in joint representations of the GUT gauge group. We discuss how comparing predictive supersymmetric GUT models with the experimental results for quark and charged lepton masses leads to constraints on the SUSY spectrum. We show an example from a recent analysis where the resulting superpartner masses where found just beyond the reach of LHC run 1, but fully within the reach of a 100 TeV pp collider.

%\keywords{Keyword1; keyword2; keyword3.}
\end{abstract}

%\ccode{PACS numbers:}

%\tableofcontents

\section{Introduction}	

Supersymmetry (SUSY) has various attractive features. Most prominent among them are the properties that SUSY ameliorates considerably the hierarchy problem of the SM, by introducing new particles, superpartners of the SM states, with spin that differs from that of the SM counterparts by half a unit. Furthermore, these new states modify the renormlization group (RG) running of the gauge couplings in a way that simple schemes for Grand Unification of the fundamental interactions become possible, with a unification scale high enough to be consistent with bounds on proton decay.

Observations tell us that supersymmetry has to be broken, such that the masses of the additional superpartner particles are (in general) heavier than the electroweak (EW) scale. From a bottom-up perspective, and from the theory point of view, the scale(s) where these masses lie is essentially a free parameter of the respective SUSY extension of the SM. On the other hand, Grand Unified Theories (GUTs) have the potential to constrain these scales, as we recently investigated in Ref.\ [\refcite{Antusch:2015nwi}], and as we will discuss in this note.

The mass scale(s) of the SUSY particles (= sparticles) is relevant for various reasons. To start with, at the LHC various searches for them have been performed with negative results. As a general rule, the lighter the sparticles the better the solution to the hierarchy problem. Because of this many people were hoping for an early discovery of SUSY at the LHC, which did not happen (so far). However, the measure of the ``severeness'' of the hierarchy problem is not possible without some ambiguity.

On the other hand, currently envisioned future 100 TeV pp colliders such as the FCC-hh and the SppC could probe SUSY particles up to mass scales of {\cal O}(10 TeV) \cite{Cohen:2013xda,Cohen:2014hxa,Ellis:2015xba}. 
In this note we like to discuss an example where a GUT scenario predicts a SUSY spectrum which may be fully testable at the FCC-hh or SppC. We will also go through the general arguments behind this result and discuss how it may generalise to other GUT models.

\section{Predictive GUTs for quark and lepton mass ratios}

GUTs are defined as theories which unify the three forces of the SM into a single unified force, described by a single gauge symmetry group. As a consequence, also the particles get unified into joint representations of this gauge group. This can lead to predictions for the ratios of quark and lepton masses, respectively their Yukawa couplings, at high energies where the GUT description holds. Which predictions for the ratios are realised depends on the model, however there is only a limited number of options. GUT models which feature such predictions for the quark-lepton Yukawa coupling ratios are much more predictive than models without this property, and are thus of high interest in the theoretical community.

One prominent example is so-called bottom-$\tau$ unification, or top-bottom-$\tau$ unification, i.e.\ the possible prediction that the respective third family Yukawa couplings are equal at the GUT scale. For the second generation, Georgi and Jarlskog postulated the GUT scale ratio $y_\mu/y_s = 3$ for the strange quark Yukawa coupling and the Yukawa coupling of the muon \cite{GJ}. More recently, driven for example by the changed experimental results for the mass of the strange quark, alternative ratios have been proposed \cite{Antusch:2009gu,Antusch:2013rxa}, for example in $SU(5)$ GUTs $y_\tau = \pm \frac{3}{2} y_b$, $y_\mu = 6 y_s$ or $y_\mu = \frac{9}{2} y_s$, and  $y_e = - \frac{1}{2} y_d$. 

To compare the GUT predictions for the quark-lepton Yukawa ratios, which hold at high energies, with the experimental results for quark and lepton masses at low energies, one has to calculate their RG running. At the scale of the SUSY particles, the sparticles have to be integrated out of the theory and the SUSY extension has to be matched to the SM at loop level. It is known that these SUSY threshold corrections\cite{threshold} can have a large effect on the Yukawa couplings, especially when they are enhanced by a large (or moderate) value of $\tan \beta$. 

The crucial point here is that these SUSY threshold corrections depend on the sparticle spectrum, i.e.\ on the masses of the SUSY particles. All the sets of GUT predictions known to date for the quark-lepton Yukawa ratios for all three families require a certain size of the threshold corrections, i.e.\ impose specific constraints on the SUSY spectrum. As we are going to show in an example, these requirements, combined with the measured value of the SM-like Higgs mass, can be powerful enough to constrain the sparticle spectrum to a range accessible by future 100 TeV pp colliders.\cite{Antusch:2015nwi}

\section{GUTs and the boundary conditions for the SUSY parameters at the GUT scale}

The fact that GUTs also unify the SM particles (and their superpartners) in joint representations of the GUT symmetry group also reduces the number of free SUSY parameters at the GUT scale. In SU(5) GUTs, for example, one is left with only two soft breaking mass matrices at the GUT scale per family, one for the fermions in the five-dimensional matter representation and one for the fermions in the ten-dimensional representation. In SO(10) GUTs, there is only one unified sfermion mass matrix. 

In addition, the symmetries of GUT flavour models like \cite{Antusch:2013kna,Antusch:2013tta,Antusch:2014poa,Gehrlein:2014wda} include various (non-Abelian) ``family symmetries'', which lead to hierarchical Yukawa matrices and impose (partially) universal soft breaking mass matrices among different generations. The combination of these effects can indeed lead to GUT scale boundary conditions which are very ``universal'' and can be described by only a few parameters. 

Furthermore, universal boundary conditions may also be a result of a specific SUSY breaking mechanism. In the example to be presented below, for simplification, we will assume Constrained MSSM (CMSSM) boundary conditions for the soft breaking parameters at the GUT scale, which is a quite strong assumption that will probably often be relaxed in realistic models. 

Finally, we like to note that the absence of deviations from the SM in flavour physics processes implies constraints on flavour non-universalities in the SUSY spectrum (if the sparticles are not too heavy) and provides an experimental hint that, if SUSY exists at a comparatively low scale, it should be close to flavour-universal. In any case, it will be interesting to investigate in future works how the constraints on the SUSY spectrum get modified when the assumption of exact CMSSM boundary conditions at the GUT scale is relaxed.

\section{Example: SUSY spectrum from GUT scenarios with $y_\tau = \pm \frac{3}{2} y_b$, $y_\mu = 6 y_s$ and  $y_e = \frac{1}{2} y_d$}

\begin{figure}
\includegraphics[scale=.41]{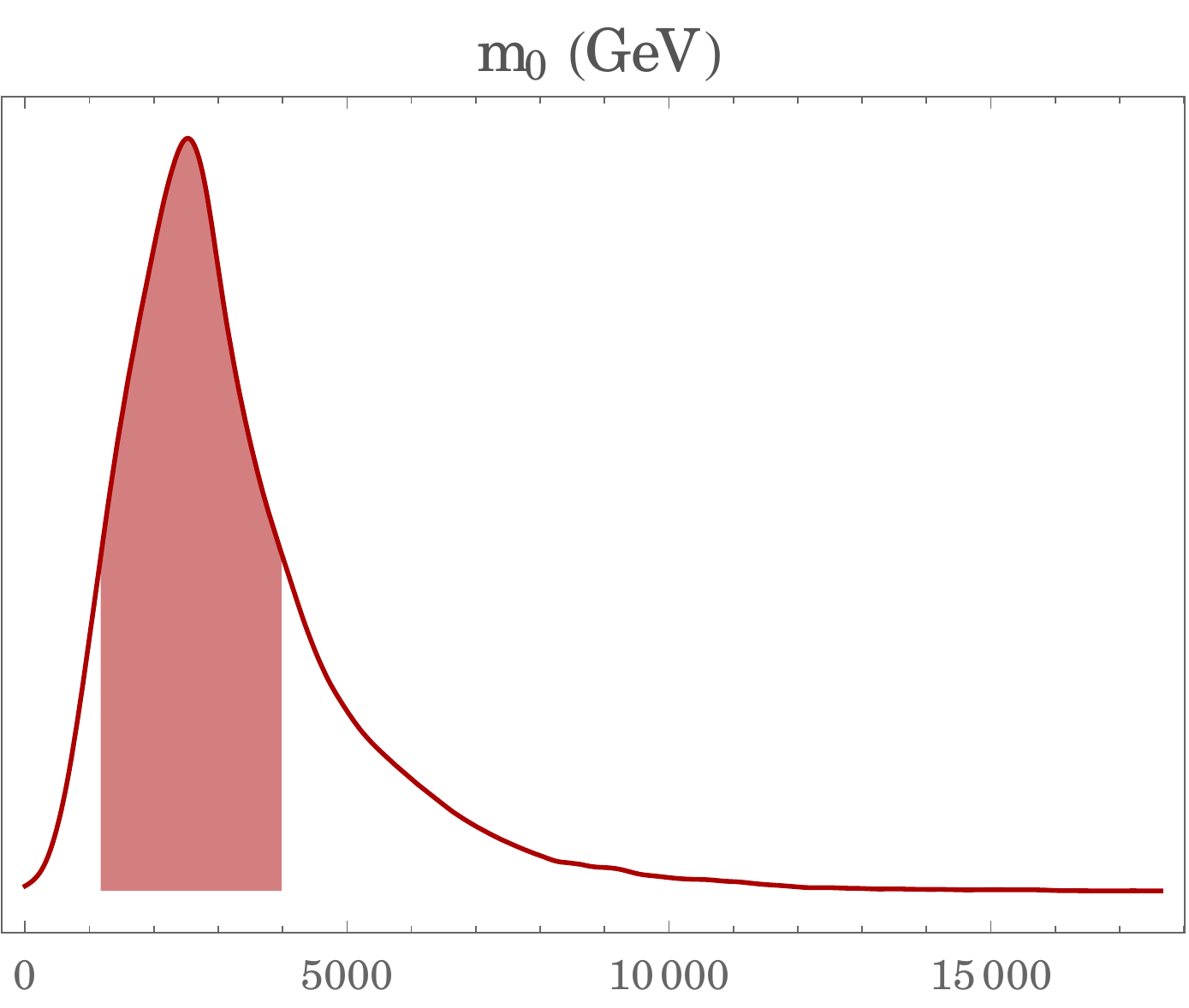}
\quad
\includegraphics[scale=.41]{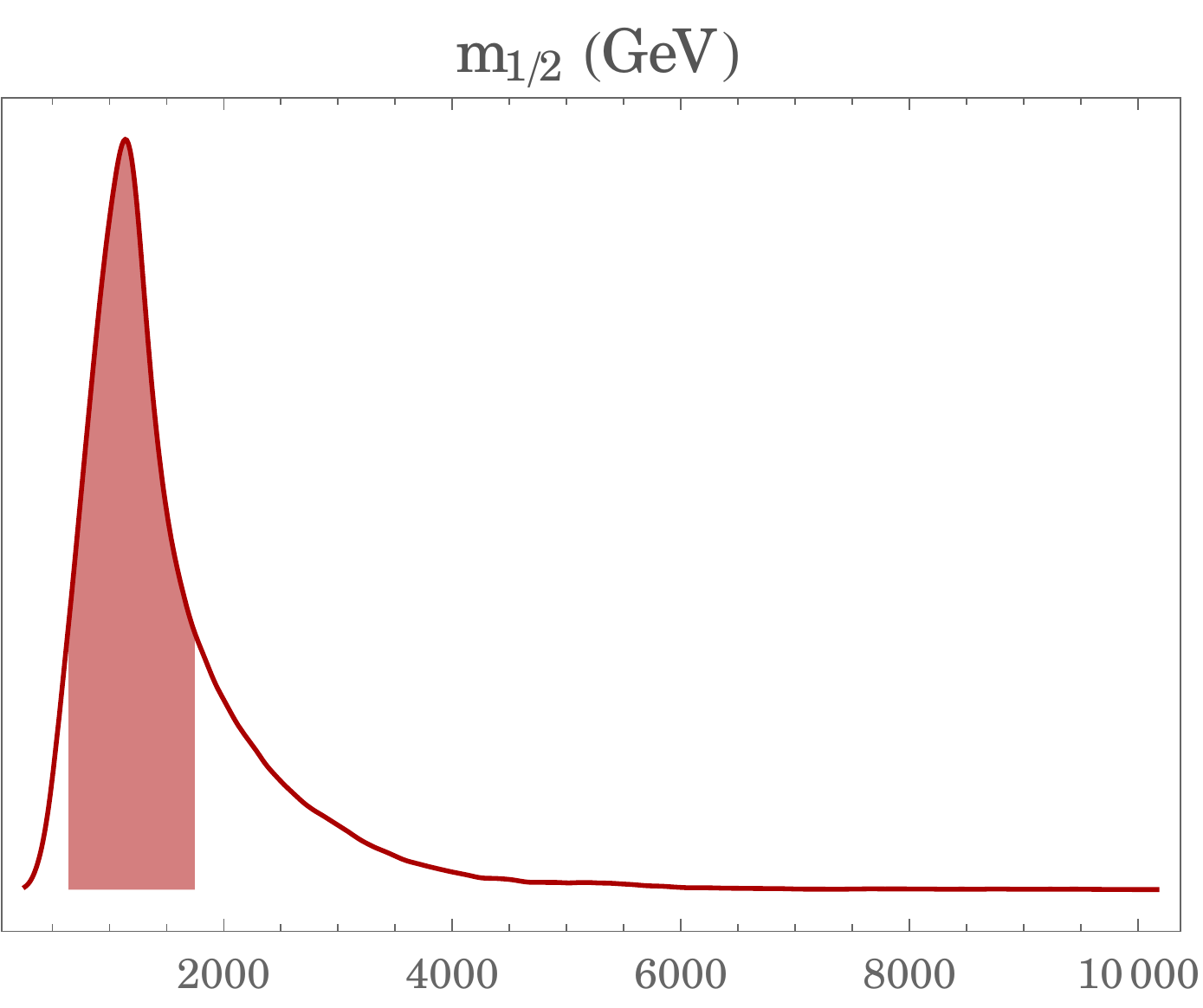}
\quad
\begin{center}
\includegraphics[scale=.41]{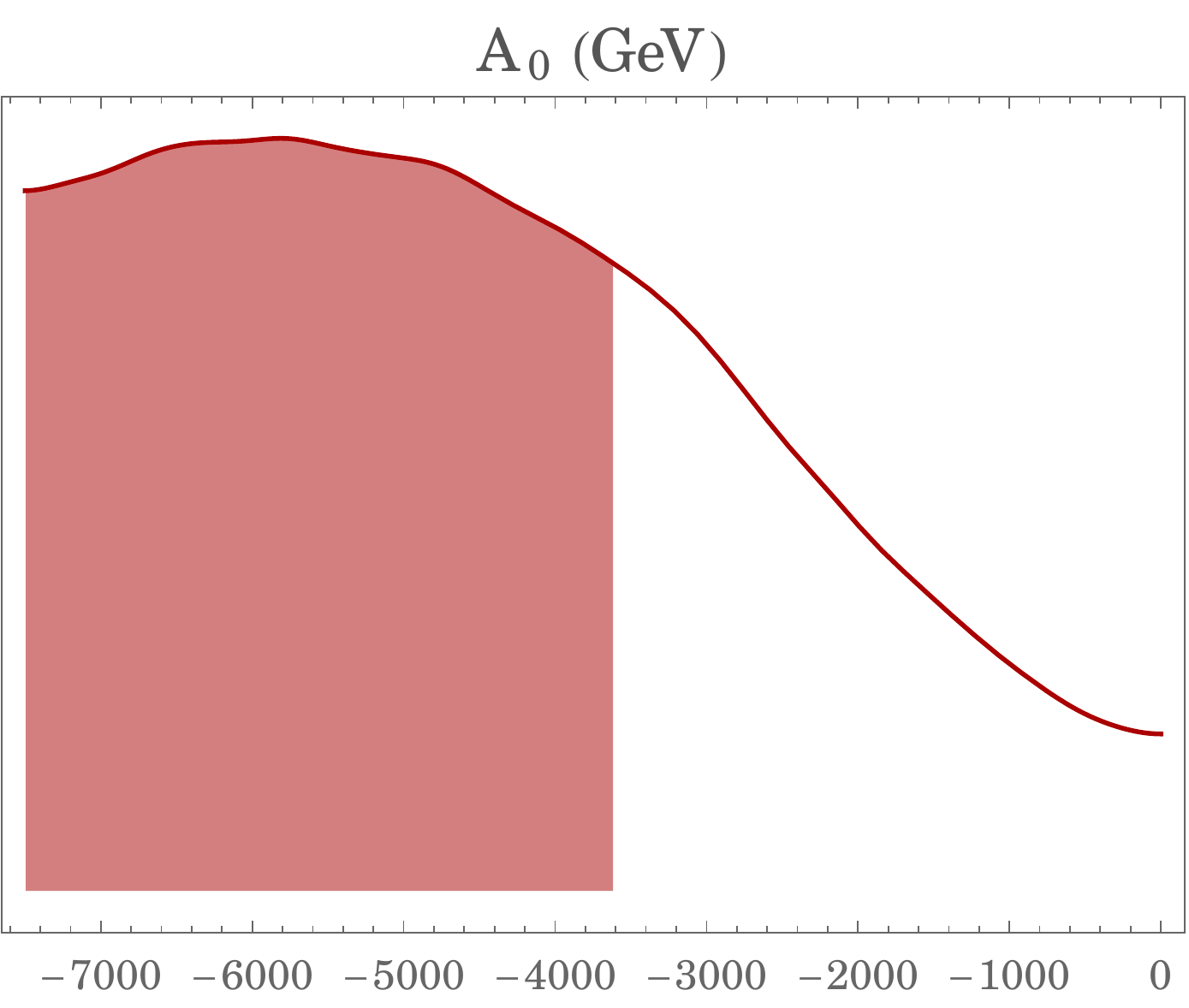}
\end{center}
\caption{$1\sigma$ HPD intervals for the Constrained MSSM soft-breaking parameters.\cite{Antusch:2015nwi}}
\label{fig:hpdcmssm}
\end{figure}

As an example, we will consider the class of GUT models which features the GUT-scale Yukawa relations $\frac{y_e}{y_d}=-\frac{1}{2}$, $\frac{y_\mu}{y_s}=6$, and $\frac{y_\tau}{y_b}=-\frac{3}{2}$ (cf.\ Ref.\ [\refcite{Antusch:2009gu}]). These GUT relations have been proven promising for GUT flavour model building and can emerge as direct result of CG factors in SU(5) GUTs or as approximate relation after diagonalization of the GUT-scale Yukawa matrices $Y_d$ and $Y_e$ (cf.\ Refs.\ [\refcite{Antusch:2013kna,Antusch:2013tta,Antusch:2014poa,Gehrlein:2014wda}]). 

For the GUT scale boundary conditions for the soft-breaking parameters we restrict our analysis to the Constrained MSSM, with parameters $m_0$, $m_{1/2}$, and $A_0$, with $\mu$ determined from requiring the breaking of electroweak symmetry, and set {\it sgn} $(\mu)=+1$. We have not included $\tan\beta$ explicitly in the fit, however we have scanned over various different values of $\tan\beta$ and found that the best fit can be obtained for values of $\tan\beta \approx 30$. We have therefore set $\tan\beta = 30$ for our main analysis. The RG running including the calculation of the SUSY threshold corrections for all families has been performed with the \texttt{REAP}\cite{Antusch:2003kp} extension \texttt{SusyTC}\cite{Antusch:2015nwi}. 

\begin{figure}
\includegraphics[scale=.45]{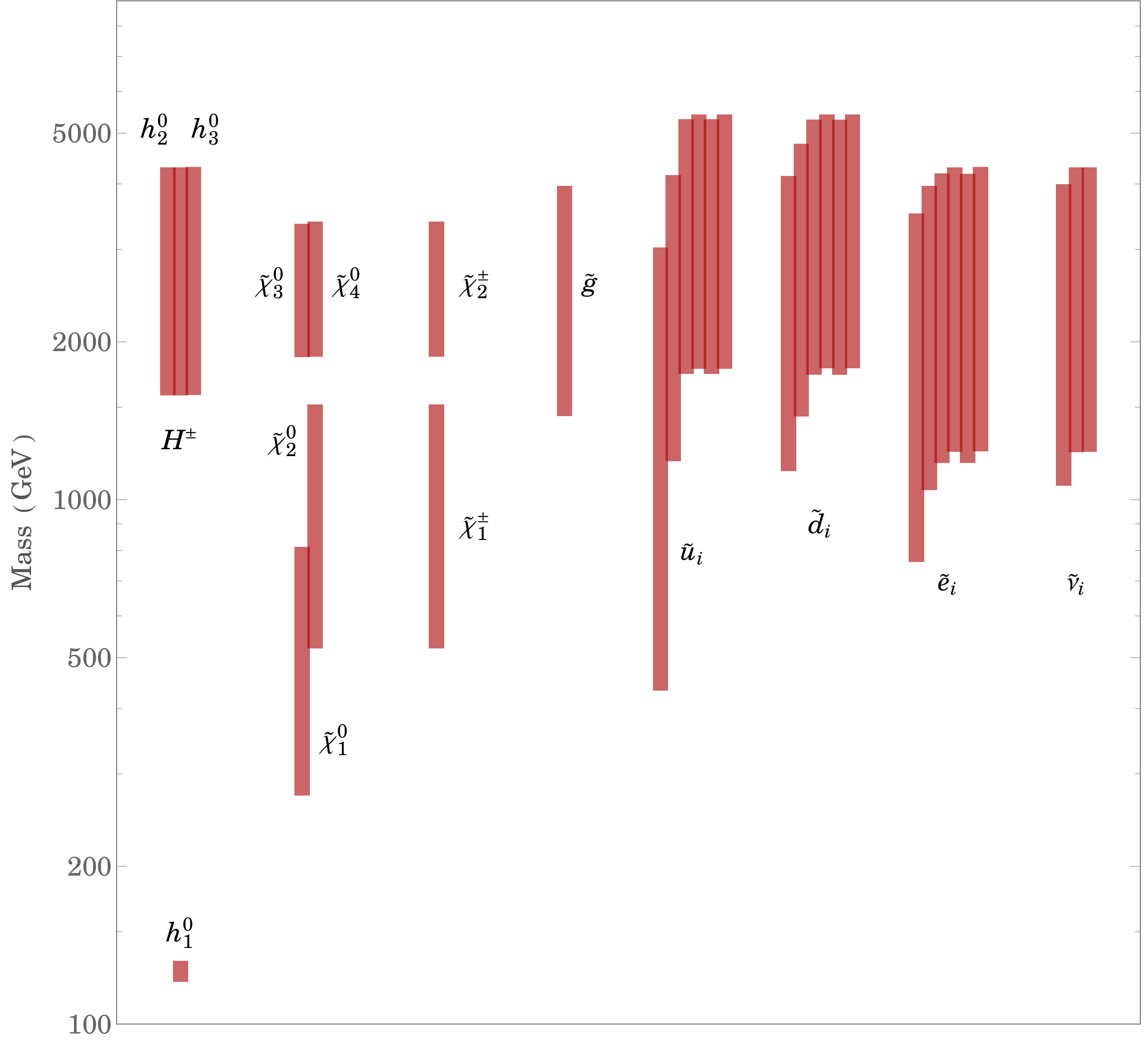}
\caption{$1\sigma$ HPD intervals for the sparticle spectrum and Higgs boson masses with $SU(5)$ GUT scale boundary conditions $\frac{y_e}{y_d}=-\frac{1}{2}$, $\frac{y_\mu}{y_s}=6$, and $\frac{y_\tau}{y_b}=-\frac{3}{2}$. The LSP is always $\tilde \chi_1^0$ and the NLSP is always a stop.\cite{Antusch:2015nwi}}
\label{fig:hpd}
\end{figure}

We use the experimental constraints for the running $\overline{\text{MS}}$ Yukawa couplings at the Z-boson mass scale calculated in Ref.\ [\refcite{Antusch:2013jca}], and set the uncertainty of the charged lepton Yukawa couplings to 1\% to account for the estimated theoretical uncertainty (which exceeds here the experimental uncertainty). When applying the measured Higgs mass $m_H=125.7\pm 0.4~\text{GeV}$ \cite{Agashe:2014kda} as constraint, we use a 1$\sigma$ interval of $\pm 3$ GeV, including the estimated theoretical uncertainty. For calculating $m_H$ at the two-loop level we have used the external software package \texttt{FeynHiggs} 2.11.2,\cite{Heinemeyer:1998yj} the current version when our numerical analysis was performed.

The confidence intervals for the masses of the sparticles are obtained as Bayesian ``highest posterior density'' (HPD) intervals from a Markov Chain Monte Carlo sample of two million points, using a Metropolis algorithm. As an additional constraint we restricted $|A_0|<7.5$ TeV to make sure to avoid too large vacuum decay rates. It would be desirable to compute the lifetime of the vacuum for each point of the Markov Chain, however this clearly would take too much computation time.  We remark that a more accurate inclusion of the lifetime constraint in the MC analysis may somewhat enlarge the predicted ranges for the masses of the sparticles. Our results for the $1\sigma$ intervals for the Constrained MSSM parameters are shown in figure \ref{fig:hpdcmssm}. The $1\sigma$ HPD results of the sparticle masses are presented in figure \ref{fig:hpd}. For all parameter points the LSP and NLSP are a neutralino and stop, respectively. The interval for the SUSY scale is $Q_\text{HPD} = [841,3092]$ GeV.

We note that the analysed GUT scale relation $\frac{y_\tau}{y_b}=-\frac{3}{2}$, $\frac{y_\mu}{y_s}=6$ and $\frac{y_e}{y_d}=-\frac{1}{2}$ is indeed only one of the possible predictions that can arise from GUTs. We have chosen the above set of GUT scale predictions since they are among the ones recently used successfully in GUT model building\cite{Antusch:2013kna,Antusch:2013tta,Antusch:2014poa,Gehrlein:2014wda}. In the future, it will of course be interesting to also test other combinations of promising GUT relations, and compare the resulting predictions for the SUSY spectra. Further details, comments and discussion of this analysis can be found in Ref.\ [\refcite{Antusch:2015nwi}].

\section{General arguments}
Although we have analyzed here a specific example only, some of the effects that lead to a predicted sparticle spectrum seem rather general, as long as the quark-lepton Yukawa ratios are predicted at the GUT scale together with (close-to) universal soft-breaking parameters:  
\begin{itemize}

\item The main reason for the predictions/constraints on the SUSY spectrum is the fact that, to our knowledge, all the possible sets of GUT predictions for the quark-lepton Yukawa ratios require a certain amount of SUSY threshold corrections for each generation.\footnote{Note that with a CMSSM-like spectrum the SUSY threshold corrections are very similar for the first two families, and therefore the argument is also valid even if the quark-lepton Yukawa ratios are predicted for two of the families only, i.e.\ for the third family and either the second or the first family.} In general, to obtain the required size of the threshold corrections, one cannot have a sparticle spectrum which is too ``split'' (as e.g.\ in Ref.\ [\refcite{splitsusy}]), since otherwise the loop functions (cf.\ Ref.\ [\refcite{Antusch:2015nwi}]) get too suppressed. More specifically, the required threshold corrections constrain the ratios of trilinear couplings, gaugino masses, $\mu$ and sfermion masses. In a CMSSM-like scenario, this implies that the ratios between $m_0$, $m_{1/2}$ and $A_0$ are constrained. Furthermore, since the most relevant threshold corrections are the ones which are $\tan \beta$-enhanced, it also implies that $\tan \beta$ cannot bee too small. 

\item With the ratios between $m_0$, $m_{1/2}$, and $A_0$ constrained and a moderate to large value of $\tan \beta$, the measured value of the mass $m_h$ of the SM-like Higgs allows to constrain the SUSY scale. We emphasise that this is an important ingredient, since the threshold corrections themselves depend only on the ratios of trilinear couplings, gaugino masses, $\mu$ and sfermion masses, and do not constrain the overall scale of the soft breaking parameters. The combination of the two effects results in a predicted sparticle spectrum from the assumed GUT boundary conditions.

\end{itemize}

Since the Higgs mass $m_h$ plays an important role, we would like to remark that it would be highly desirable to have a more precise computation of $m_h$ available, especially for the ``large stop-mixing'' regime. In our analysis, we have used a theoretical uncertainty of $\pm 3$ GeV, which is dominating the 1$\sigma$ interval for $m_h$. This theoretical uncertainty should of course, strictly speaking, not be treated on the same footing as a pure statistical uncertainty. Furthermore, there are indications that the theoretical uncertainty in the $m_h$ calculation in the most relevant regions of parameter space of our analysis, with ``large stop-mixing'', may be larger (as  recently discussed\cite{workshop}), however there is no full agreement on this. For our example analysis, as mentioned above, we have used the external software \texttt{FeynHiggs} 2.11.2\cite{Heinemeyer:1998yj} for a two-loop calculation of the Higgs mass, the current version when our numerical analysis was performed, and the most commonly assumed estimate $\pm 3$ GeV for the theoretical uncertainty.

\section{Summary}
We have discussed how certain classes of predictive GUT models are capable of predicting a testable SUSY spectrum at a future 100 TeV pp collider such as the FCC-hh or the SppC. The predictions for the sparticle spectrum can be understood as follows:

When GUT models predict the ratios of quark and charged lepton masses for all three generations at the GUT scale, as a result of the unification of the SM particles in GUT representation, they impose constraints on the amount of SUSY threshold corrections. This in turn implies constraints on the SUSY spectrum. These constraints, combined with the measured value of the SM-like Higgs mass, can be powerful enough to constrain the sparticle spectrum to a compact region.\cite{Antusch:2015nwi}

We have discussed an example where we found (cf.\ figure \ref{fig:hpd}) that the resulting superpartner masses are beyond the reach of LHC run 1, but fully within the reach of a 100 TeV pp collider.

\subsection*{Acknowledgements}
This work was supported by the Swiss National Science Foundation. We thank the organisers of the IAS Program on the Future of High Energy Physics in Hong Kong for their hospitality.

\bibliographystyle{unsrt}

\begin{thebibliography}{}


\bibitem{Antusch:2015nwi}
  S.~Antusch and C.~Sluka,
  %``Predicting the Sparticle Spectrum from GUTs via SUSY Threshold Corrections with SusyTC,''
  arXiv:1512.06727 [hep-ph].
  
  
\bibitem{Cohen:2013xda}
  T.~Cohen, T.~Golling, M.~Hance, A.~Henrichs, K.~Howe, J.~Loyal, S.~Padhi and J.~G.~Wacker,
  %``SUSY Simplified Models at 14, 33, and 100 TeV Proton Colliders,''
  JHEP {\bf 1404} (2014) 117
%  doi:10.1007/JHEP04(2014)117
  [arXiv:1311.6480 [hep-ph]].

\bibitem{Cohen:2014hxa}
  T.~Cohen, R.~T.~D'Agnolo, M.~Hance, H.~K.~Lou and J.~G.~Wacker,
  %``Boosting Stop Searches with a 100 TeV Proton Collider,''
  JHEP {\bf 1411} (2014) 021
  %doi:10.1007/JHEP11(2014)021
  [arXiv:1406.4512 [hep-ph]].

\bibitem{Ellis:2015xba}
  S.~A.~R.~Ellis and B.~Zheng,
  %``Reaching for squarks and gauginos at a 100 TeV p-p collider,''
  Phys.\ Rev.\ D {\bf 92} (2015) no.7,  075034
 % doi:10.1103/PhysRevD.92.075034
  [arXiv:1506.02644 [hep-ph]].

  \bibitem{GJ}
  H.~Georgi and C.~Jarlskog,
  %``A New Lepton - Quark Mass Relation In A Unified Theory,''
  Phys.\ Lett.\  B {\bf 86} (1979) 297.
  %%CITATION = PHLTA,B86,297;%%
  
   %\cite{Antusch:2009gu}
\bibitem{Antusch:2009gu}
  S.~Antusch and M.~Spinrath,
  %``New GUT predictions for quark and lepton mass ratios confronted with phenomenology,''
  Phys.\ Rev.\ D {\bf 79} (2009) 095004
  [arXiv:0902.4644 [hep-ph]].
  %%CITATION = ARXIV:0902.4644;%%

  %\cite{Antusch:2013rxa}
\bibitem{Antusch:2013rxa}
  S.~Antusch, S.~F.~King and M.~Spinrath,
  %``GUT predictions for quark-lepton Yukawa coupling ratios with messenger masses from non-singlets,''
  Phys.\ Rev.\ D {\bf 89} (2014),  055027
%  doi:10.1103/PhysRevD.89.055027
  [arXiv:1311.0877 [hep-ph]].
  %%CITATION = doi:10.1103/PhysRevD.89.055027;%%

 \bibitem{threshold}
 %\cite{Hempfling:1993kv}
%\bibitem{Hempfling:1993kv}
  R.~Hempfling,
  %``Yukawa coupling unification with supersymmetric threshold corrections,''
  Phys.\ Rev.\ D {\bf 49} (1994) 6168, 
  %%CITATION = PHRVA,D49,6168;%%
 %\cite{Hall:1993gn}
%\bibitem{Hall:1993gn}
  L.~J.~Hall, R.~Rattazzi and U.~Sarid,
  %``The Top quark mass in supersymmetric SO(10) unification,''
  Phys.\ Rev.\ D {\bf 50} (1994) 7048
  %doi:10.1103/PhysRevD.50.7048
  [hep-ph/9306309],    
  %\cite{Carena:1994bv}
%\bibitem{Carena:1994bv}
  M.~Carena, M.~Olechowski, S.~Pokorski and C.~E.~M.~Wagner,
  %``Electroweak symmetry breaking and bottom - top Yukawa unification,''
  Nucl.\ Phys.\ B {\bf 426} (1994) 269
  [hep-ph/9402253];
  %%CITATION = HEP-PH/9402253;%%
  %\cite{Blazek:1995nv}
  T.~Blazek, S.~Raby and S.~Pokorski,
  %``Finite supersymmetric threshold corrections to CKM matrix elements in the large tan Beta regime,''
  Phys.\ Rev.\ D {\bf 52} (1995) 4151
  [hep-ph/9504364], 
  %%CITATION = HEP-PH/9504364;%%
    S.~Antusch and M.~Spinrath,
  %``Quark and lepton masses at the GUT scale including SUSY threshold corrections,''
  Phys.\ Rev.\ D {\bf 78} (2008) 075020
  [arXiv:0804.0717 [hep-ph]].
  %%CITATION = ARXIV:0804.0717;%%










   %\cite{Antusch:2012fb}
\bibitem{Antusch:2012fb}
  S.~Antusch, C.~Gross, V.~Maurer and C.~Sluka,
  %``\theta^PMNS_13 = \theta_C / \sqrt2 from GUTs,''
  Nucl.\ Phys.\ B {\bf 866} (2013) 255
  [arXiv:1205.1051 [hep-ph]].
  
  %\cite{Antusch:2013kna}
\bibitem{Antusch:2013kna} 
  S.~Antusch, C.~Gross, V.~Maurer and C.~Sluka,
  %``A flavour GUT model with $\theta_{13}^{PMNS} \simeq \theta_C/\sqrt2$,''
  Nucl.\ Phys.\ B {\bf 877}, 772 (2013)
  [arXiv:1305.6612 [hep-ph]].
  %%CITATION = ARXIV:1305.6612;%%
  
    %\cite{Antusch:2013tta}
\bibitem{Antusch:2013tta} 
  S.~Antusch, C.~Gross, V.~Maurer and C.~Sluka,
  %``Inverse neutrino mass hierarchy in a flavour GUT model,''
  Nucl.\ Phys.\ B {\bf 879}, 19 (2014)
  [arXiv:1306.3984 [hep-ph]].
  %%CITATION = ARXIV:1306.3984;%%

   %\cite{Antusch:2014poa}
\bibitem{Antusch:2014poa}
  S.~Antusch, I.~de Medeiros Varzielas, V.~Maurer, C.~Sluka and M.~Spinrath,
  %``Towards predictive flavour models in SUSY SU(5) GUTs with doublet-triplet splitting,''
  JHEP {\bf 1409} (2014) 141
  [arXiv:1405.6962 [hep-ph]].
  %%CITATION = ARXIV:1405.6962;%%

  \bibitem{Gehrlein:2014wda}
  J.~Gehrlein, J.~P.~Oppermann, D.~Schäfer and M.~Spinrath,
  %``An SU(5) $\times$ A$_5$ golden ratio flavour model,''
  Nucl.\ Phys.\ B {\bf 890} (2014) 539
  doi:10.1016/j.nuclphysb.2014.11.023
  [arXiv:1410.2057 [hep-ph]].

\bibitem{Antusch:2003kp}
  S.~Antusch, J.~Kersten, M.~Lindner and M.~Ratz,
  %``Running neutrino masses, mixings and CP phases: Analytical results and phenomenological consequences,''
  Nucl.\ Phys.\ B {\bf 674} (2003) 401
%  doi:10.1016/j.nuclphysb.2003.09.050
  [hep-ph/0305273].

    %\cite{Antusch:2013jca}
\bibitem{Antusch:2013jca}
  S.~Antusch and V.~Maurer,
  %``Running quark and lepton parameters at various scales,''
  JHEP {\bf 1311} (2013) 115
  [arXiv:1306.6879 [hep-ph]].
  %%CITATION = ARXIV:1306.6879;%%

\bibitem{Agashe:2014kda}
  K.~A.~Olive {\it et al.} [Particle Data Group Collaboration],
  %``Review of Particle Physics,''
  Chin.\ Phys.\ C {\bf 38} (2014) 090001.
  
\bibitem{Heinemeyer:1998yj}
  S.~Heinemeyer, W.~Hollik and G.~Weiglein,
  %``FeynHiggs: A Program for the calculation of the masses of the neutral CP even Higgs bosons in the MSSM,''
  Comput.\ Phys.\ Commun.\  {\bf 124} (2000) 76
  [hep-ph/9812320];
  %%CITATION = HEP-PH/9812320;%%
  %\cite{Heinemeyer:1998np}
%\bibitem{Heinemeyer:1998np}
  S.~Heinemeyer, W.~Hollik and G.~Weiglein,
  %``The Masses of the neutral CP - even Higgs bosons in the MSSM: Accurate analysis at the two loop level,''
  Eur.\ Phys.\ J.\ C {\bf 9} (1999) 343
  [hep-ph/9812472];
  %%CITATION = HEP-PH/9812472;%%
  %\cite{Degrassi:2002fi}
%\bibitem{Degrassi:2002fi}
  G.~Degrassi, S.~Heinemeyer, W.~Hollik, P.~Slavich and G.~Weiglein,
  %``Towards high precision predictions for the MSSM Higgs sector,''
  Eur.\ Phys.\ J.\ C {\bf 28} (2003) 133
  [hep-ph/0212020];
  %%CITATION = HEP-PH/0212020;%%
  %\cite{Frank:2006yh}
%\bibitem{Frank:2006yh}
  M.~Frank, T.~Hahn, S.~Heinemeyer, W.~Hollik, H.~Rzehak and G.~Weiglein,
  %``The Higgs Boson Masses and Mixings of the Complex MSSM in the Feynman-Diagrammatic Approach,''
  JHEP {\bf 0702} (2007) 047
  [hep-ph/0611326];
  %%CITATION = HEP-PH/0611326;%%
  %\cite{Hahn:2013ria}
%\bibitem{Hahn:2013ria}
  T.~Hahn, S.~Heinemeyer, W.~Hollik, H.~Rzehak and G.~Weiglein,
  %``High-Precision Predictions for the Light CP -Even Higgs Boson Mass of the Minimal Supersymmetric Standard Model,''
  Phys.\ Rev.\ Lett.\  {\bf 112} (2014),  141801
  [arXiv:1312.4937 [hep-ph]].
  %%CITATION = ARXIV:1312.4937;%%

  \bibitem{splitsusy} 
%\cite{ArkaniHamed:2004fb}
%\bibitem{ArkaniHamed:2004fb}
  N.~Arkani-Hamed and S.~Dimopoulos,
  %``Supersymmetric unification without low energy supersymmetry and signatures for fine-tuning at the LHC,''
  JHEP {\bf 0506} (2005) 073
  [hep-th/0405159];
  %\cite{Giudice:2004tc}
%\bibitem{Giudice:2004tc}
  G.~F.~Giudice and A.~Romanino,
  %``Split supersymmetry,''
  Nucl.\ Phys.\ B {\bf 699} (2004) 65
  [hep-ph/0406088].

\bibitem{workshop}
 Workshop ``Precision SUSY Higgs Mass Calculation Initiative'', Heidelberg, 20-22.01.2016, https://sites.google.com/site/kutsmh/home.


\end{thebibliography}

\end{document}